\documentclass[%
 reprint,
 amsmath,amssymb,
 aps,
]{revtex4-2}
\usepackage{graphicx}
\usepackage{dcolumn}
\usepackage{bm}
\usepackage[usenames]{color}
\usepackage{colortbl}
\usepackage{hyperref}

\begin{document}

\title{Lateral Josephson Junctions as Sensors for Magnetic Microscopy at Nano-Scale}

\author{Razmik A. Hovhannisyan$^{1,2}$}
\author{Sergey Yu. Grebenchuk$^{1,3}$}%
\author{Denis S. Baranov$^{1,4,5}$}
\author{Dimitri Roditchev$^{6,7}$}
\author{Vasily S. Stolyarov$^{1,4,8}$}
\email{vasiliy.stoliarov@gmail.com}
\affiliation{$^1$ Advanced mesoscience and nanotechnology centre, Moscow Institute of Physics and Technology, Dolgoprudny 141700, Russia}
\affiliation{$^2$ Department
of Physics, Stockholm University, AlbaNova University Center,
SE-10691 Stockholm, Sweden} 
\affiliation{$^3$ Department of Materials Science and Engineering, National University of Singapore, 117575, Singapore, Singapore
}
\affiliation{$^4$ Dukhov Research Institute of Automatics (VNIIA), 127055 Moscow, Russia
}
\affiliation{$^5$ National Research Nuclear University MEPhI (Moscow Engineering Physics Institute), 115409 Moscow, Russia
}
\affiliation{$^6$ LPEM UMR-8213, ESPCI Paris, PSL Research University, CNRS, 75005 Paris, France}
\affiliation{$^7$ INSP UMR-7588, Sorbonne Universite, CNRS, 75005 Paris, France}
\affiliation{$^8$National University of Science and Technology MISIS, 119049 Moscow, Russia\\}

\begin{abstract}
Lateral Josephson junctions (LJJ) made of two superconducting Nb electrodes coupled by Cu-film are applied to quantify the stray magnetic field of Co-coated cantilevers used in Magnetic Force Microscopy (MFM). The interaction of the magnetic cantilever with LJJ is reflected in the electronic response of LJJ as well as in the phase shift of cantilever oscillations, simultaneously measured. The phenomenon is theorized and used to establish the spatial map of the stray field. Based on our findings, we suggest integrating LJJs directly on the tips of cantilevers and using them as nano-sensors of local magnetic fields in Scanning Probe Microscopes. Such probes are less invasive than conventional magnetic MFM cantilevers and simpler to realize than SQUID-on-tip sensors. \cite{Finkler2010local}.
\end{abstract}

\date{Accepted for publication in the Journal of Physical Chemistry Letters, 10 December 2021}
\maketitle

\preprint{AIP/123-QED}
\date{\today}


\section{Introduction}
Invented a long ago~\cite{barone1982physics}, lateral Josephson junctions (LJJ) are key elements of superconducting quantum electronic devices ~\cite{soloviev2021miniaturization}: single photon detectors~\cite{chen2011microwave,walsh2017graphene,murch20121}, thermometers~\cite{faivre2014josephson}, RF-sensors~\cite{russer2011nanoelectronic}, among many others. LJJ are also widely used in fundamental studies~\cite{boris2013evidence,likharev1986dynamics,hoss2000multiple} as they are extremely sensitive to magnetic field that creates superconducting phase gradients, redistributes supercurrents, causes Josephson vortex entry~\cite{roditchev2015direct,dremov2019local}, thus modifying the electronic response of LJJ~\cite{tinkham1996introduction,wallraff2003quantum,embon2017imaging,rowell1963magnetic,golod2021reconfigurable}.




The lateral geometry of LJJ makes them suitable for studying Josephson phenomena at the nanoscale by Scanning Probe Microscopies (SPM) such as  Tunneling Microscopy and Spectroscopy~\cite{roditchev2015direct,serrier2013scanning,cherkez2014proximity,Niu_2013,kim2012visualization,stolyarov2020electronic,yano2021magnetic}, Scanning SQUID~\cite{vasyukov2013scanning,kirtley2016scanning,anahory2020squid,bouchiat2009detection,Zeldov_2017} or Magnetic Force Microscopy (MFM)~\cite{polshyn2019manipulating,dremov2019local,grebenchuk2020observation,Serri_2017}. The latter technique is particularly promising owing to its simplicity and its capacity to reveal the spatial distribution of the magnetic field over LJJ. Yet quantitative MFM measurements are rare, due to difficulties related to the calibration of magnetic MFM probes. To overcome the issue, several works used magnetic objects with well-known geometry and magnetic properties, such as nanowires~\cite{kebe2004calibration}, ferromagnetic stripes~\cite{vock2010monopolelike}, mono-dispersive magnetite nanoparticles~\cite{sievers2012quantitative}. In Ref.~\cite{di2019quantitative}, it has been shown how quantized flux of Abrikosov vortex and antivortex pair can be used as an object for the calibration procedure. The main issue with these approaches is that they rely on the measurement of the interaction force which depends on stray magnetic field of the probe but also on several other parameters~\cite{carneiro2000vortex}, difficult to reproduce.

\begin{figure}[!ht]
\begin{center}
\includegraphics[width=0.8\columnwidth]{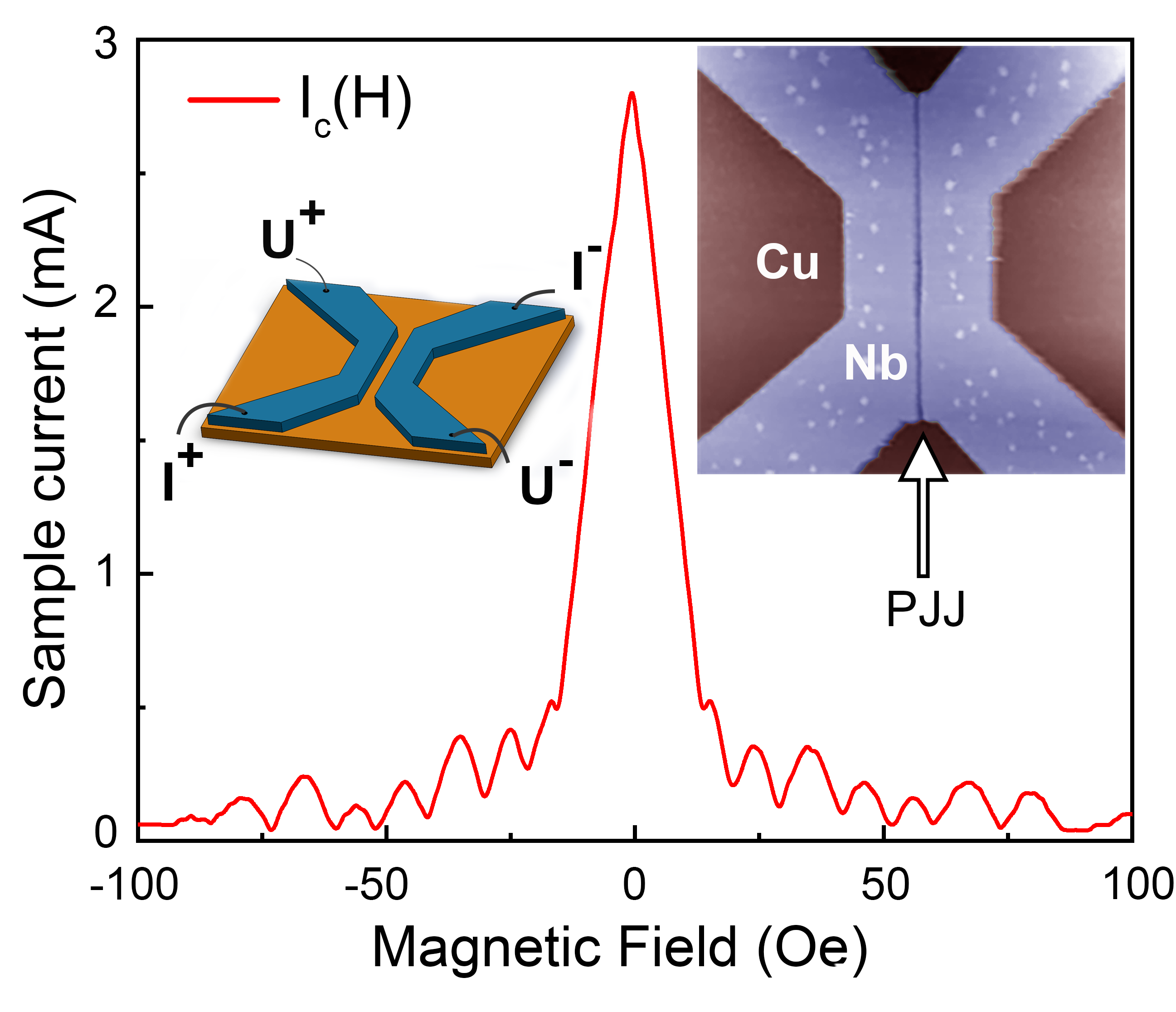}
\caption{Magnetic field dependence of the critical current of the used LJJ device. $I_c(H_{ext})$ is symmetric with respect to the direction of the magnetic field $H_{ext}$ applied perpendicularly to the device plane. A linear decrease at low fields is typical for long Josephson junctions. The oscillations reflect penetration of individual Josephson vortices. Left inset: schematic illustration of studied device. I$^+$, I$^-$ and U$^+$, U$^-$ denote current and voltage electrodes, respectively. Right inset: 3$\times$3 $\mu$m$^2$ AFM image of the device. White arrow points to the Cu-gap between the two Nb-electrodes. 
}
\label{fig:scheme}
\end{center}
\end{figure}

In the present work, Nb/Cu/Nb LJJ  were built to calibrate a standard magnetic  Co/Cr-coated cantilever~\cite{MESP} by using superconducting flux quanta of Josephson vortices as a calibrating unit.The left inset in Fig.~\ref{fig:scheme} schematizes the investigated device, the right inset is the top view AFM image of it. The details of the LJJ fabrication and measurement procedure can be found in Appendix Materials.
Furthermore, by studying the transport characteristics of LJJ when moving the MFM probe above the active area of the junction, we determine the spatial distribution of the stray magnetic field generated by cantilever. Interestingly, our measurements demonstrate the possibility to use LJJ as magnetic field sensors offering a high spatial resolution. We finally design a local probe based on LJJ built directly at the tip apex of the cantilever. This opens a route for reliable quantitative measurements of magnetic properties at the nanoscale by SPM.  

\section{Results and Discussion}
Fig.~\ref{fig:scheme}  shows the variation of the critical Josephson current $I_c$ of the device as a function of the externally applied perpendicular magnetic field $H_{ext}$ (red curve). The left inset details the used four-terminal measurement scheme. $I_c(H_{ext})$ shows oscillations and has a triangular dependence around zero-field, typical for long Josephson junctions~\cite{owen1967vortex}. 

\begin{figure}
    \begin{center}
    \includegraphics[width=0.95\columnwidth]{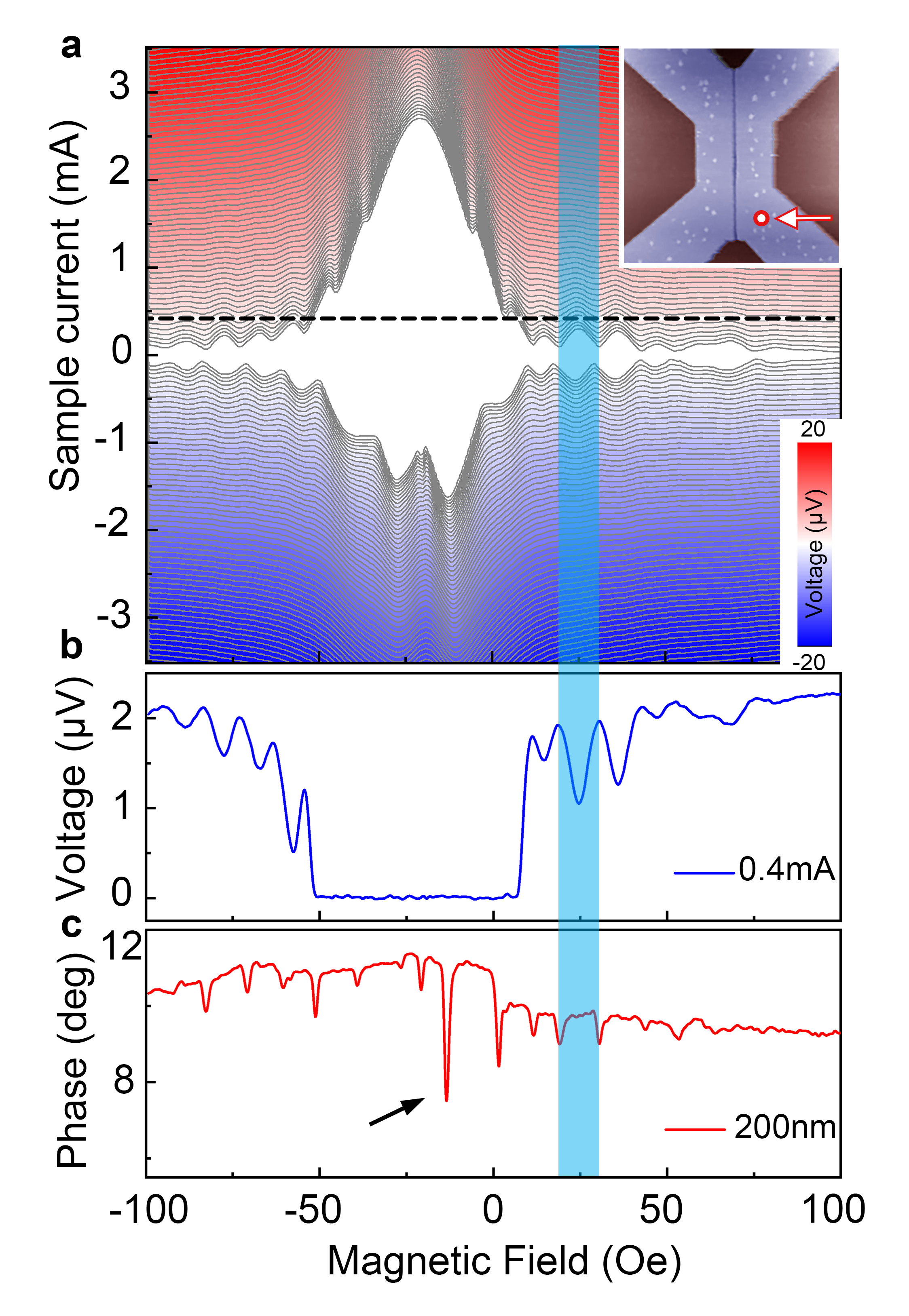}
    \caption{Interaction of the device with MFM cantilever. (a) -  color-coded $V(I,H_{ext})$ characteristics of the device with the cantilever situated above the device at $h$= 200~nm. The lateral position of the cantilever is shown in the inset by arrow and white dot. The presence of cantilever induces strong distortions and asymmetry in $I_c(H_{ext})$ with respect to both $I$ and $H_{ext}$, as compared to intrinsic behavior (thin black lines, from Fig.~\ref{fig:scheme}). (b) - magnetic field dependence of the voltage drop $V$ across the device at bias currents $I=$ 0.4 mA (corresponding to black horizontal dashed line  (a). (c) - magnetic field dependence of the cantilever oscillation phase (measured at zero-bias current). The vertical blue band underlines the match between the oscillations of $I_c(H_{ext})$ in (a), $V(H_{ext})$ in (b) and phase vs $H_{ext}$ in (c), all capturing the 4-th Josephson vortex penetration in the device.}
    \label{fig:tip-position}
    \end{center}
\end{figure}
\begin{figure}
    \begin{center}
    \includegraphics[width=0.9 \columnwidth]{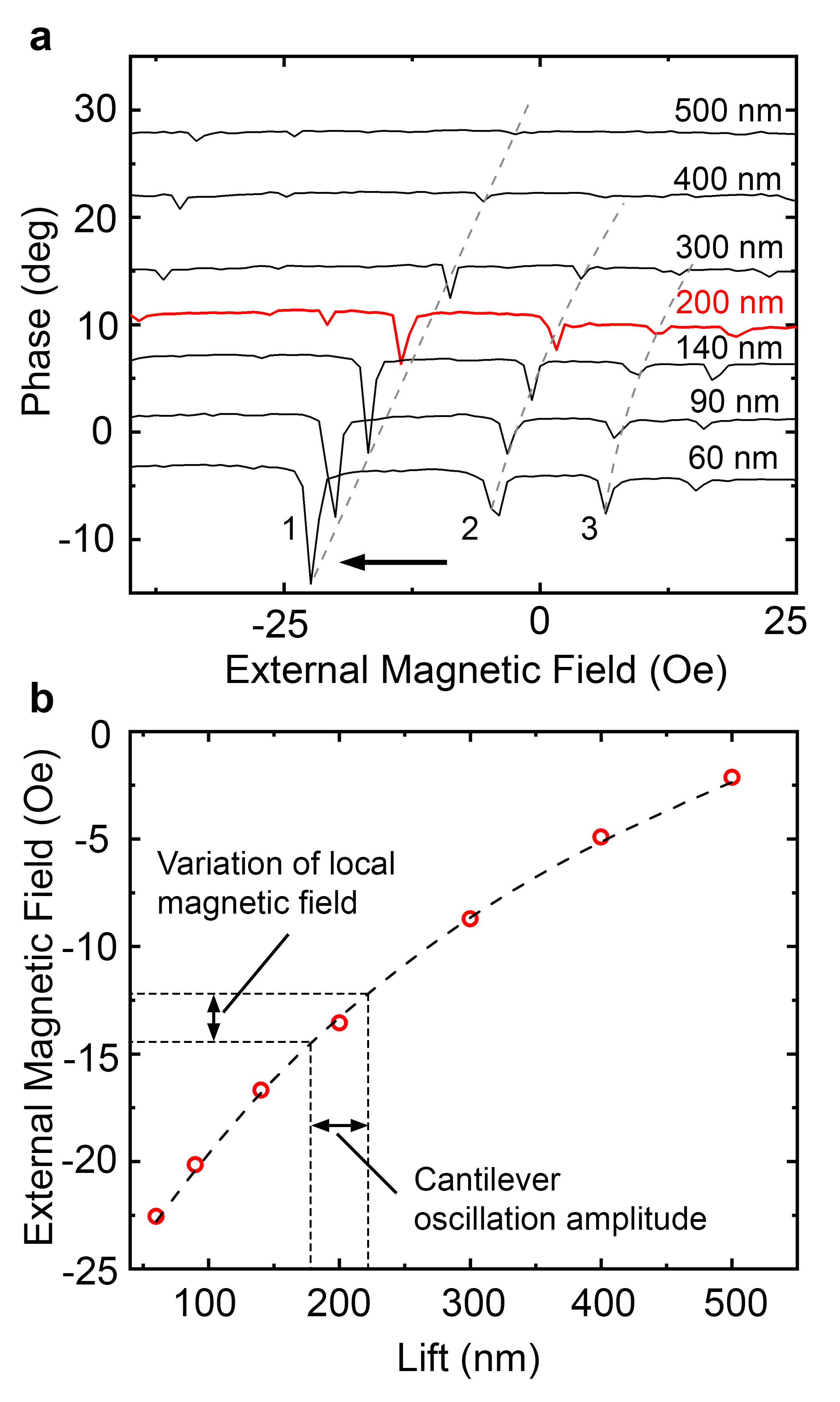}
    \caption{Cantilever stray field determination. (a) - oscillation phase versus external magnetic field at different heights $h$ (lifts) of the cantilever above the device. Sharp drops correspond to the penetration of individual Josephson vortices into the junction. Grey dashed lines link the drops corresponding to the entry of 1st, 2nd and 3rd Josephson vortices. Black arrow points on the series of drops due to the penetration of the 1st Josephson vortex occurring when the sum of the external and cantilever's fields matches one flux quantum inside the junction. (b) - external field of 1st Josephson vortex penetration as a function of cantilever lift. Red circles - data points taken from (a); black dashed line - best fit (see in the text). Vertical and horizontal dashed lines enable visualizing the relationship between the oscillation amplitude (20 nm) and resulted variations of the magnetic field.} 
    \label{fig:phase-fit}
    \end{center}
\end{figure}

 Approaching the MFM cantilever to the active area of the device leads to dramatic modifications of its magnetic response~\cite{samokhvalov2012properties,dremov2019local}. In the experiment presented in Fig.~\ref{fig:tip-position}, the cantilever was positioned 200 nm above one of the Nb contacts of the device (the lateral position of the cantilever is shown in the inset by a white dot). The color-coded $V(I,~H_{ext})$ map in Fig.~\ref{fig:tip-position}(a) shows that despite a large cantilever-device distance, not only the maximum of the Fraunhofer-like pattern shifts away from zero-field but also the pattern itself  becomes deformed and strongly asymmetric with respect to the direction of the applied current~\cite{dremov2019local,krasnov2020josephson}. 

 The cut  $V(H_{ext})$ of this map taken at fixed $I=$ 0.4 mA is plotted in Fig.~\ref{fig:tip-position}(b). Other cuts are presented in supplementary Fig.~S2. There, the voltage variations have the same origin as critical current oscillations in (a)~\cite{tinkham1996introduction} revealing the AC-Josephson phenomenon: each voltage drop between two neighbouring maxima corresponds to a given integer number of Josephson vortices inside LJJ device. As the external or cantilever-induced field is modified, the number of Josephson vortices changes. This provides an opportunity to detect the total flux created by stray field of the cantilever with an accuracy reaching a small fraction of the magnetic flux quantum $\Phi_0$. 

 
To validate the approach, an additional experiment was provided in which the cantilever was left in the same position (inset in Fig.~\ref{fig:tip-position}(a)) and the phase of its oscillation was registered as a function of the external field, Fig~\ref{fig:tip-position}(c). On this curve, each drop is caused by dissipation during one Josephson vortex entry~\cite{dremov2019local}.
For instance, the black arrow points to the phase drop occurring in the moment of the penetration of the first Josephson vortex into the junction. The physical phenomenon behind this drop is an abrupt variation of the repulsive force acting on cantilever due to significant redistribution of Meissner currents in LJJ device upon Josephson vortex entry. The vertical blue rectangle in Figs. ~\ref{fig:tip-position}(a-c) delimits the field range corresponding to 4 Josephson vortices inside the LJJ. Note, that on $V(H_{ext})$ curve in Fig~\ref{fig:tip-position}(b), the entry of the first and second Josephson vortex is not detected because at this bias current the device remains in the superconducting state (full set of $V(H_{ext})$ recorded at different $I$ is presented in Supplementary Fig.~S2; the spatial maps of $dV/dI(x,y)$  - in Supplementary Fig.~S4). Though, these events are unambiguously detected in $I_c(H_{ext})$, Fig~\ref{fig:tip-position}(a), and in phase vs $H_{ext}$, Fig~\ref{fig:tip-position}(c). 

The peculiar magnetic field response of LJJ device presented in ~Fig.~\ref{fig:tip-position} is further used for cantilever calibration. For this, the cantilever is positioned at different heights $h$ above the device, and the magnetic field dependence of the oscillation phase is recorded; the results are presented in~Fig.~\ref{fig:phase-fit}(a). It is straightforward to see that the intensity of the external magnetic field at which Josephson vortex penetrate depend on the height of the cantilever above the device. The main effect is the shift of the phase drops, as showed by black arrows in Fig.~\ref{fig:phase-fit}(a) pointing to the first Josephson vortex penetration. This shift is related to the  variation of the magnetic contribution $\Phi_{tip}$ of the cantilever to the total magnetic flux penetrating the device, $\Phi_0$ at the first Josephson vortex penetration,

\begin{equation}
    \begin{split}
        \Phi_0  = \Phi_{ext}(H_{ext}) + \Phi_{tip}(h),\\
    \end{split}
\end{equation}
 where $\Phi_{ext}$ is the flux created by external field.  
 
 To calculate $\Phi_{tip}$, we modelled the stray field of the cantilever by a magnetic charge $q$ situated at a distance $\delta$ from the tip apex~\cite{di2019quantitative} (see section Appendix B). The results of calculations are presented in Fig~\ref{fig:phase-fit}(b). There, red open circles show the field of the first Josephson vortex penetration plotted as a function of the distance $h$ between the cantilever and LJJ device; these data are extracted directly from Fig~\ref{fig:phase-fit}(a). The dashed line is the best fit using the model. The fitting parameters, $q = 6.6\cdot10^{-8}$~Am ($ 8.25\cdot10^{-14}$~Wb) and $\delta = 301.5$ nm, are in a good agreement with the results obtained in Ref.~\cite{di2019quantitative}. 
 
 While the fit was made for a steady field, the oscillation of the cantilever (20 nm amplitude in the present case) leads to an oscillating contribution $\Phi_{tip}$ in the total flux penetrating the device.  Dashed lines in Fig~\ref{fig:phase-fit}(b) enable one to appreciate the importance of this effect. As can be seen, it leads to the field modulation in the range of $H_f = 2-3$ Oe which is significantly smaller than the field period $\sim$10 Oe  of $I_c(H_{ext})$ oscillations, thus enabling distinguishing separate vortex states. In general, the resolution of the presented approach depends on the size of the LJJ and the amplitude of cantilever oscillation. The latter should eventually be adjusted/limited to enable distinguishing different vortex states inside the device.  
 
 \begin{figure}[!ht]
\begin{center}
\includegraphics[width=0.95\columnwidth]{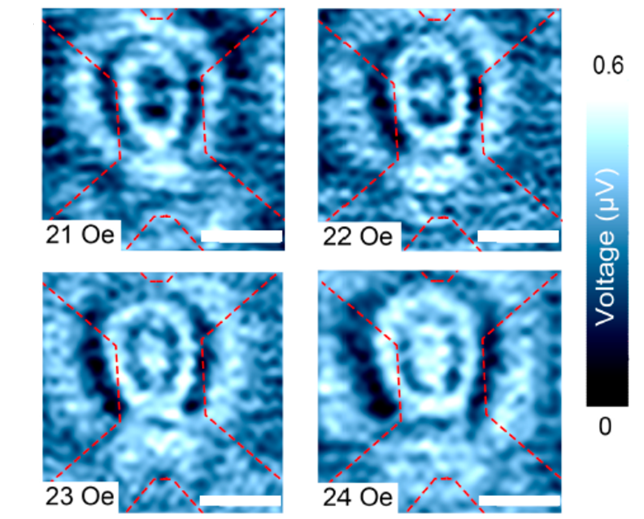}
\caption{Imaging Josephson vortex penetration. Spatial maps of the voltage drop $V$ across the device ($I=$ 0.8 mA) acquired at external fields 21 Oe, 22 Oe, 23 Oe and 24 Oe with the MFM cantilever moving laterally above the device center over 3$\times$3$~\mu$m$^2$ area (lift -- 200 nm). The white scale bar corresponds to 1 $\mu m$. Red dashed lines show the location of the device. Concentric rings with increasing radii delimit the regions with, respectively, 4, 3, and 2 Josephson vortices inside the device. The ring sizes continuously increase with the external field. }
\label{fig:volt}
\end{center}
\end{figure}

  \begin{figure}[!ht]
\begin{center}
\includegraphics[width=0.95\columnwidth]{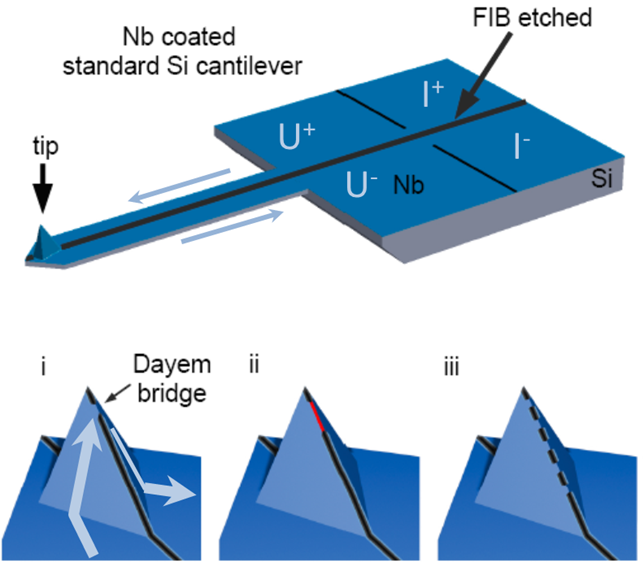}
\caption{Proposed LJJ-based probes. The probes comprise an insulating AFM cantilever (in grey) covered with a thin superconducting layer (here Nb, in dark blue) on an eventual metallic sub-layer. The probes are patterned by FIB to remove superconducting parts (black regions). 4 contacts on the probe base are used for measuring $V(I)$ characteristics. Light blue arrows show the supercurrent flow in the device. Close to the tip apex, see (i), the supercurrent flows through LJJ. Different LJJ geometries are suggested: i - short Dayem bridge LJJ; ii - long LJJ (metallic link is shown in red) and iii - multi-hole sensor built on the tip of the cantilever (see in the text).}
\label{fig:probe}
\end{center}
\end{figure}

 In some quantitative MFM experiments the only knowledge of $\Phi_{tip}(h)$ is not enough, and the full data-set $\Phi_{tip}(x,y,h)$ is necessary. Unfortunately, each MFM cantilever has its individual geometry, and this dependence is not universal. However, it is possible to use LJJ device to establish it rather precisely. To do this, the cantilever is kept at a fixed height $h$ and is laterally scanned above the active area of LJJ device. The total flux varies, resulting in variations of the voltage across LJJ device measured at a fixed $I$. The results of such measurements are presented in Fig. ~\ref{fig:volt} where the color-coded map of the voltage drop $V(x,y,h=200$~nm) measured at $I$ = 0.8 mA) is presented for various intensities of the external field. On these maps, the voltage oscillations form bright and dark concentric circles. By varying the external field, the size of the circles is modified. The origin of the phenomenon is identical to that of the phase drops recorded in MFM measurements, Figs.~\ref{fig:tip-position}(c),~\ref{fig:phase-fit}(a). In fact, the voltage maxima (bright areas in Fig.~\ref{fig:volt}) correspond to the entry/exit of a new Josephson vortex in the LJJ; the area limited by two neighbouring circles of voltage maxima corresponds to a fixed number of Josephson vortices in the junction. 
 
 Thus, bright (but also dark) circles at $V = const$ represent the contours of known iso-flux. Such maps can be used to calibrate MFM tips. Moreover, by combining these maps with the magnetic charge modeling (see Appendix B), enables obtaining a full three-dimensional map of the tip stray field with a high spatial $\sim$20-100 nm and field $\sim$1 Oe resolution (see Appendix C). 
 
The $V(x,y,h)$ maps demonstrate a high sensitivity of LJJ to local magnetic field. This leads us to suggest a new probe  for scanning probe microscopy with the design based on Josephson junctions patterned directly on the tip apex, Fig.~\ref{fig:probe}. A thin layer (in blue) of a superconducting material (Nb, Al, NbN,..) or a S/N sandwich (Nb/Cu, Nb/Ti,..) is deposited onto an insulating cantilever (in grey). The superconducting film is then patterned: Some conducting parts are removed using a focused ion beam (black stripes)\cite{levichev2019proximity}. The minimalistic design comprises a Dayem bridge put close to the tip apex (i). The very apex remains free thus enabling simultaneous AFM experiments; also, it protects LJJ from damages caused by accidental crashes of the tip on the studied sample. The excitation and $V(I)$ readout are done using four electrodes I$^+$, I$^-$, U$^+$, U$^-$ patterned on the base of the cantilever. Other geometries are easily achievable, such as a long LJJ (ii) or a multi SQUID device (iii). In some applications, these designs advantageously differ from SQUID-on-tip approaches by their simplicity of realization, variability of shapes, behaviour in high magnetic fields \footnote{During the manuscript preparation, we became aware about the work~\cite{wyss2021magnetic} suggesting to integrate a SQUID on the AFM cantilever}. The Dayem bridge design suggested in Fig.~\ref{fig:probe}(i) has another important advantage: it has a very low amount of superconducting material located close to the studied sample and thus, by its poor diamagnetic screening, it is potentially less invasive than a SQUID-on-tip sensor \cite{wyss2021magnetic}. Into the opposite, a long LJJ could be advantageous in cases when an increased field sensitivity due to field focusing effect in LJJ~\cite{stolyarov2020josephson} or a high spatial resolution in one direction are required~\cite{golod2019planar}.

\section{Conclusion}

To conclude, we demonstrated a quantitative method for measuring the spatial distribution of spatially varying magnetic fields at the nanoscale; the method is based on the detection of individual Josephson vortices entering/escaping lateral Josephson junctions. The fact that each Josephson vortex carries one flux quantum along with a small size lateral junctions enables using such devices as highly sensitive non-invasive detectors of the magnetic field. In the present study, we applied the method to calibrate a standard MFM cantilever. The detection was done by following the drops of the cantilever oscillation phase and, simultaneously, by measuring the voltage peaks across the junction when the first Josephson vortex penetrated into the junction. We demonstrated that our approach can be used for a very accurate determination of stray fields created by a magnetic cantilevers. Furthermore, we suggest a new sensor design based on Josephson junctions patterned directly on the tip apex of a cantilever. This could be an alternative to the existing magnetic sensitive devices and can be used as a non-invasive sensor even for single-molecule studies.


\bibliography{aipsamp}
%
\newpage
\appendix
\renewcommand{\figurename}{Supplementary Figure}
\setcounter{figure}{0}  
\section{Methods}
Lateral Nb/Cu/Nb SNS junctions were fabricated using UHV magnetron sputtering, e-beam lithography technique with hard mask, and plasma-chemical etching as follows. First, a 50-nm Cu film and 100 nm Nb film were subsequently deposited onto SiO$_2$/Si substrate in a single vacuum cycle. The polymer mask for Nb leads was then formed by electron lithography. The pattern was covered by a 20-nm-thick aluminum layer lifted off, the Al hard mask for Nb leads was formed. Next, uncovered Nb was etched by the plasma-chemical process. After Nb patterning the Al mask was removed by wet chemistry.

The resulted LJJ device (studied in this work) has the following geometrical characteristics: the junction length is 2500 nm, its width is 200 nm, the width of Nb leads in the JJ area is 500 nm (see insets in main Fig.~1). More details on the sample fabrication and characterization can be found in \cite{dremov2019local}.

Transport measurements were performed using four-probe configuration. The critical temperature of the sample was 7.2 K at zero external magnetic field with cantilever retracted far from the sample. The maximum critical current at T=4.2K was $I_c$=2.8 mA  (see main Fig. 1).

The AFM/MFM measurements were conducted using scanning probe system (AttoCube AttoDry 1000/SU) at a temperature close to 4.2 K. The external magnetic field up to $\pm$100 Oe) was applied perpendicularly to the LJJ plane. Both AFM topography and MFM magnetic response data were obtained using  Co/Cr-coated cantilever (Bruker, MESP V2, k=2.8 N/m) oscillating at its resonance frequency, around 87 kHz. The tip oscillation amplitude and phase were measured at constant frequency with deactivated PLL.

During the acquisition of voltage maps (main Fig.~4), the voltage was measured using ADC/DAC electronics of Attodry1000 system. The measurements were performed at a constant scan speed 2 $\mu$m/s; the integration time per data point was 20 ms.

\section{Point probe approach}
A few papers have used the point-probe approach to calculate the force acting on an MFM tip due to the diamagnetic response of a superconductor to the stray field of the tip itself in the monopole approximation \cite{Reittu_1992,HUG1991357,di2019quantitative}. In the present work, to evaluate the magnetic field of the MFM cantilever, we also used the so-called point-probe approach as most simple. In the framework of this method, the stray field of our cantilever is taken in the form of a magnetic monopole Fig.~S\ref{figS}(a,b), and approximated as: 
\begin{equation*}
    \mathbf{H}_{tip} = \frac{q}{4\pi \mu_0}\frac{\mathbf{r}}{r^3},\quad(S1)
\end{equation*}
were $q$ is the magnetic charge to be determined, and $\mathbf{r}$ - the distance from the monopole. In this case, the magnetic flux created in the LJJ  by the magnetic charge of the cantilever located at a distance $z$ above it is equal to
\begin{equation*}
\begin{split}
   \Phi_{tip}(h) = \alpha\mu_0 \iint_S (\mathbf{H}_{tip} \cdot d\mathbf{S}) = \\ =  \frac{\alpha q}{4\pi }\int_{x1}^{x2}\int_{y1}^{y2}\frac{z}{r^3}dxdy  = \\ = \frac{\alpha q}{4\pi }\sum_{i= 1}^{2}\sum_{j=1}^2 (-1)^{i+j}F(x_i,y_j,h+\delta),\quad(S2)
\end{split}
\end{equation*}
where $z=h+\delta$, $h$ is the tip lift provided from AFM experiment, $\delta$ accounts for an eventual shift of the magnetic charge location with respect to the tip apex, and $\alpha$ is the geometrical factor accounting for the flux focusing effect due to Meissner diamagnetism in superconducting leads.  The $xy$-plane was chosen coinciding with LJJ surface. The function $F$ in $S2$ is the result of two-dimensional integration; it is equal to 
\begin{figure}
    \begin{center}
        \includegraphics[width=.99\columnwidth]{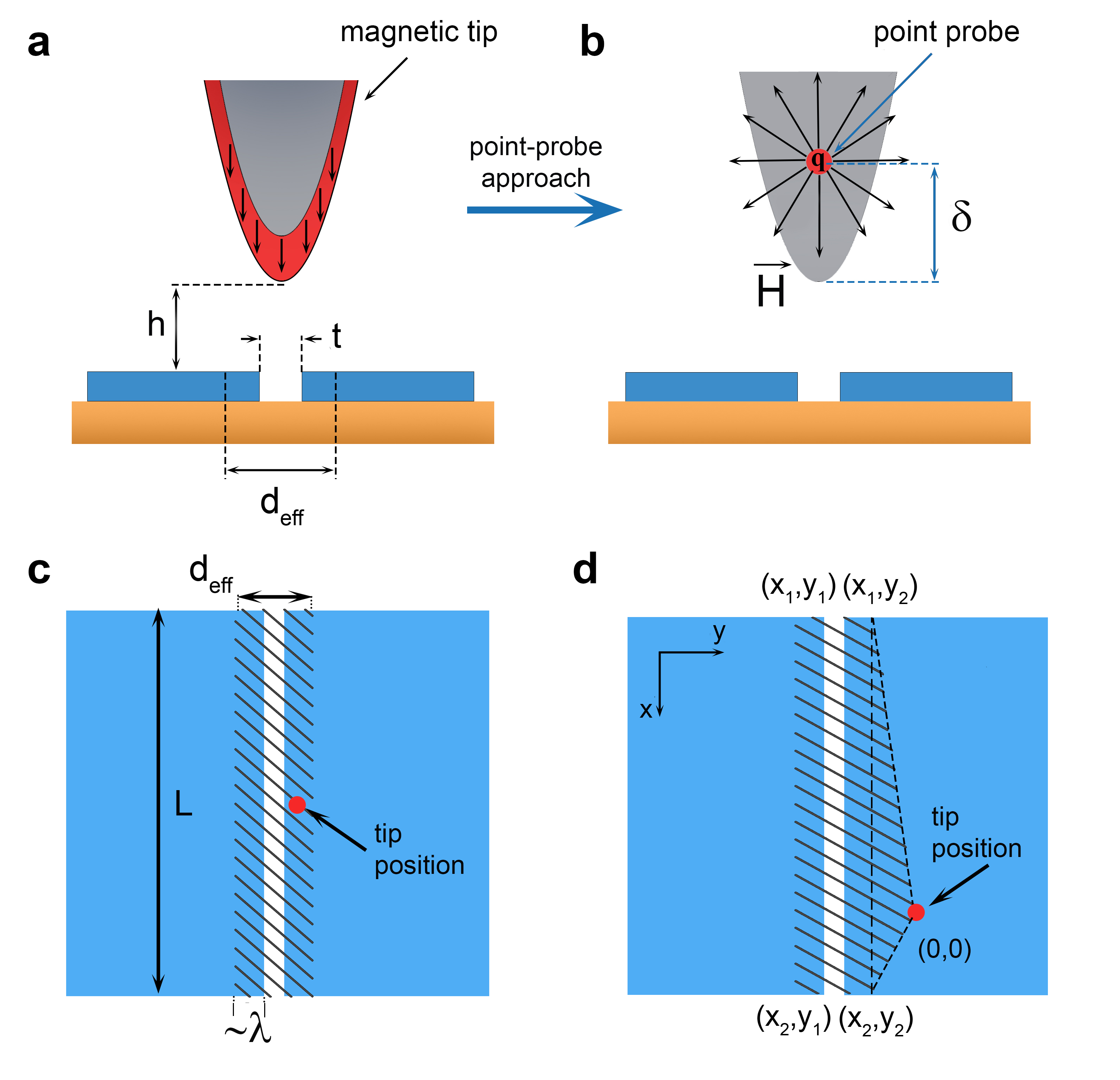}
        \caption{(a-b) - Schematic representation of the point probe approach. (a) - cantilever with magnetic coating (in red) is placed above the LJJ. The stray field of the tip is modelled by a magnetic charge $q$ situated at a distance $\delta$ from the tip apex (b). (c-d) - schematic top-view of LJJ with two superconducting electrodes S in blue and normal junction in white. The area $S =  L\times d_{eff}$ over which the integral Eq. $S2$ is taken is hatched; (c) - when the tip is located above the area $S$; (d) - when the tip is outside $S$, an additional contribution from the triangular area delimited by dashed lines is considered (see in the text).}
        \label{figS1}
    \end{center}
\end{figure}
\begin{equation*}
    \begin{split}
        F(x,y,h+\delta)= \\ = \arctan \bigg[  \frac{xy}{(h+\delta)\sqrt{x^2+y^2+(h+\delta)^2}}\bigg]\quad(S3)
    \end{split}
\end{equation*}
The surface for the flux calculation has been taken $S=L\times d_{eff}$ 
, were $L = 2500$~nm is the length of the LJJ and $d_{eff}$ is the so-called magnetic thickness of the junction (see Fig. S1(c) ). In our case, $d_{eff} = t + 2\lambda \coth(d/\lambda) \simeq 400$ nm, where $t$=200 nm is the distance between the two Nb-electrodes, $\lambda$=90 nm is the London penetration depth in Nb, and $d$=100 nm is the Nb-film thickness.

In the calculations, the integration limits were taken equal to  $x_1 = -0.14L$, $x_2 = 0.86 L$, $y_1 = 0.42d_{eff}$, $y_2 = 1.42d_{eff}$, to account for the lateral position of the tip shown by white dot in the inset of main Fig.2. The coefficient $\alpha$ was extracted from the periodicity  of measured $I_c(H_{ext})$ oscillations (shown in main Fig.1). Indeed, each oscillation corresponds to the entry in LJJ of 1 additional flux quantum $\Phi_{0}=h/(2e)$=2.068$\times$10$^{-15}$ Wb. This gives the oscillation periodicity $\Delta H=\Phi_{0}/(L\times d_{eff})\simeq$ 20 Oe with respect to the magnetic field $H$ penetrating the junction, $H=\alpha H_{ext}$. The measured periodicity $\Delta H_{ext}$= 10 $\pm$ 2 Oe is two times smaller than $\Delta H$ and therefore, $\alpha \simeq$ 2. The magnetic flux created in LJJ by external magnetic field is $\Phi_{ext} = \alpha H_{ext}\times L\times d_{eff}$; the total flux through LJJ is $\Phi_{LJJ} = \Phi_{ext} + \Phi_{tip}(h)$.

During the fitting process, the free parameters $q$ and $\delta$ were adjusted. The best fits presented in main Fig.3b were obtained with $q = 6.6\cdot10^{-8}$~Am ($ 8.25\cdot10^{-14}$~Wb) and $\delta = 301.5$ nm.

 Note, that in the case of the cantilever located outside of the area $S$, accounting for the field focusing requires an additional contribution to be added to the integral ($S4$)~\cite{golod2021reconfigurable}. It concerns the triangular area delimited in Fig.~S\ref{figS1}(c,d) by dashed lines. This contribution can be calculated using the equation 
\begin{eqnarray*}
        \tan(\Omega/2) =\\ \frac{[\mathbf{R_1},\mathbf{R_2},\mathbf{R_3}]}{R_1R_2R_3 + (\mathbf{R_1,R_2})R_3 +(\mathbf{R_1,R_3})R_2 + (\mathbf{R_3,R_2})R_1}\\
        \quad(S4)
\end{eqnarray*}
were $R_1,R_2,R_3$ are the vectors from the monopole to the vertices of the triangle and $\Omega$ is the solid angle. Then the total flux through this area is
\begin{equation*}
    \Phi_{\triangle} = \frac{q}{4 \pi} \Omega \quad (S5)
\end{equation*}

\begin{figure}
   \begin{center}
        \includegraphics[width=.85\columnwidth]{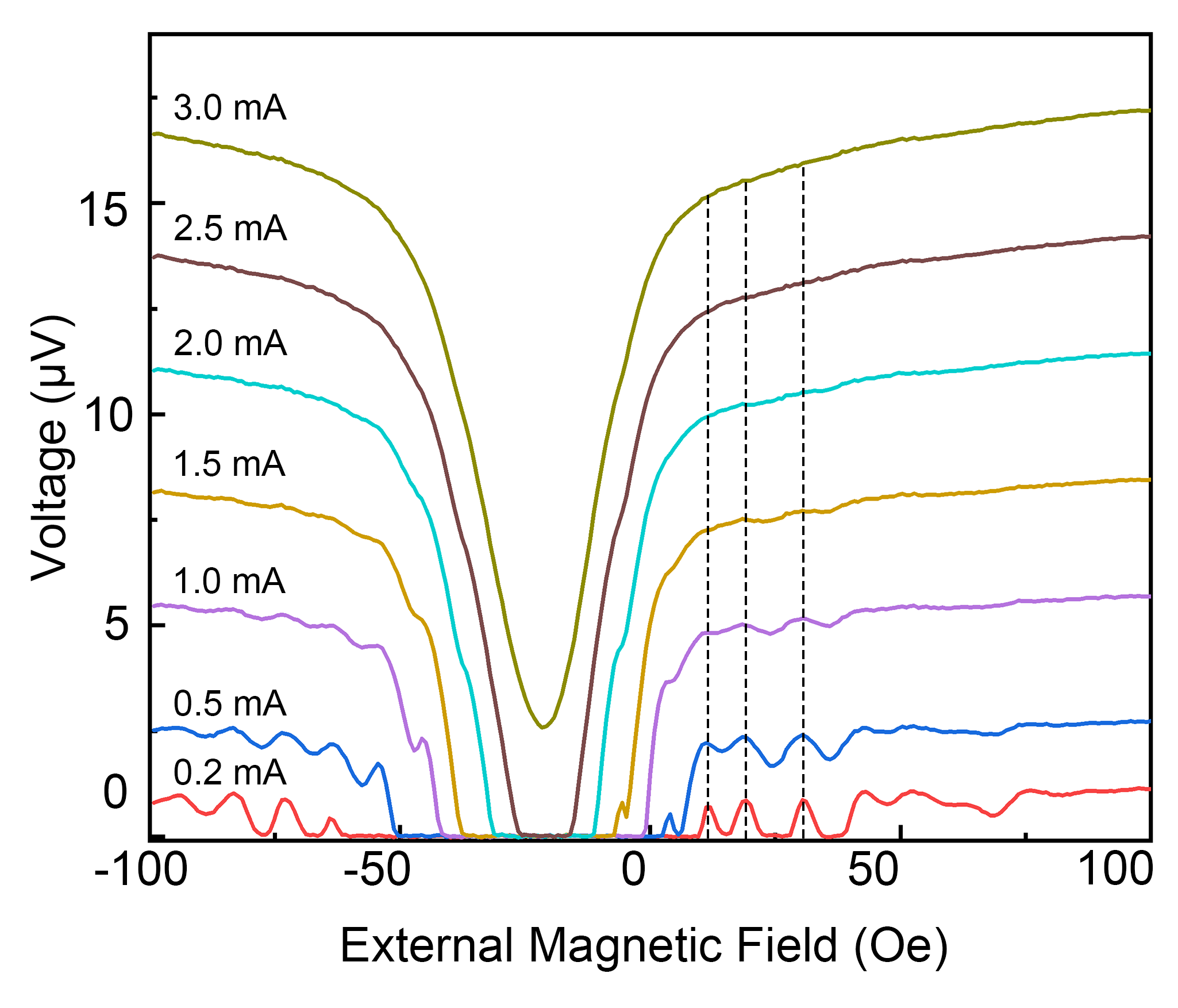}
        \caption{Magnetic field dependence of the voltage drop $V$ across the device at different bias currents. Vertical dashed lines indicate that the position of maxima of $V(H_{ext})$ oscillations does not depend on bias current.}
        \label{figS}
    \end{center}
\end{figure}

\section{Possible procedures of MFM tip calibration}

To know the amplitude of the magnetic field below the tip, one may follow the procedure used to obtain the data presented in main Fig.3. The tip is placed (at a given lift) above the center of LJJ, and the external field is adjusted to get the 1st Josephson vortex penetrate the LJJ. The moment is detected as the phase drops or as the voltage rises across the LJJ, as described in the main text and main Fig.2. At this moment, there are $N$ flux quanta inside the junction of the effective area $S$, and $H_0=N\Phi_0$/$S$ inside the junction. At the same time, $H_0$ is the sum of the tip field $H_{tip}$ ($H_{tip}$ is $z$-component of the vector field $\mathbf{H}_{tip}$) and $H_{ext}$, the latter is known. Thus, the tip field is $H_{tip}=H_0 - H_{ext}$. Note, that at this stage $S$ and $\delta$ are not known exactly. They are found by repeating the above procedure at various lifts and using the magnetic charge model. This enables to determine $H_{tip}$ at any lift (as described in the main text, Fig.3b). Moreover, by repeating this procedure at different lateral positions of the tip and using the magnetic charge model, it becomes possible to reconstruct the full 3D map of the vector magnetic field associated with it.

Another possible calibration route is based on lateral displacement of the tip over LJJ at fixed lifts. Indeed, from main Fig.1 we know that in the absence of the tip, the 1st Josephson vortex penetrates at $H_{ext}\sim$ 13 Oe, the 2nd at 19 Oe , 3rd at 30 Oe, 4th at 40 Oe, 5th at 52 Oe, 6th at 62 Oe,.., that is with the periodicity of 10 $\pm$ 2 Oe.  This periodicity depends on junction geometry and used materials but is not sensitive to temperature (well below $T_c$). The tip induces an extra field and makes additional Josephson vortices enter (or exit, if the tip field is anti-parallel to $H_{ext}$) the junction, depending on the relative position LJJ—tip. An example is shown in main Fig.4 where, at a constant lift of 200 nm, one, two and three additional Josephson vortices enter LJJ, depending on how far (laterally) the magnetic charge (tip) is situated. These events correspond to local voltage maxima that form bright concentric rings in the voltage maps. In main Fig.4, the outer bright ring corresponds to the tip positions at which the 4th Josephson vortex enters the junction (see main Figs.2b,c). Note that the entries of the 2nd and 3rd are not clearly detected as at $I$=0.8 mA the voltage contrast is too low (see main Fig.2b; a full set of $V(H_{ext})$ recorded at different $I$ is shown in supplementary Fig.S2). The intermediate bright ring corresponds to 5th vortex entry, and the inner ring – to the 6th vortex.  In main Fig.4a, $H_{ext}$=21 Oe. One can calculate the magnetic field produced in the junction by the tip situated on the outer ring (4th  vortex entry) as $H_{tip}$ = 40 Oe - $H_{ext}$ = 40-21=19 Oe. This holds for all tip locations over the outer bright ring. Note that when the external field is slightly changed (to 22 Oe in main Fig.4b), the radius of the outer ring increases. When the tip is situated on this ring, it produces the field 40-22=18 Oe. On the outer rings of Figs. 4c,d, the tip field is 17 Oe and 16 Oe respectively. For the second (intermediate) rings of main Fig.4, the 5nd vortex enters at the total field  52 Oe. At these positions, the tip creates the field 52– 21=31 Oe in Fig.4a, 30 Oe in Fig. 4b, 29 Oe in Fig.4c, and 28 Oe in Fig.4d. For the inner ring (6th Josephson vortex entry) the fields are 62-21=42 Oe, 43 Oe and 44 Oe, respectively. The resulting $H_{tip}$ map - see Supplementary Fig.S3 - is extremely precise as it enables resolving contours separated by 1 Oe in field and $\sim$(20-100) nm in space!  By repeating the procedure at different lifts, it becomes possible to reconstruct precisely the 3D distribution of the field component perpendicular to LJJ. The total (vector) field distribution can be found by fitting with magnetic charge model.

\begin{figure}
    \begin{center}
        \includegraphics[width=.9\columnwidth]{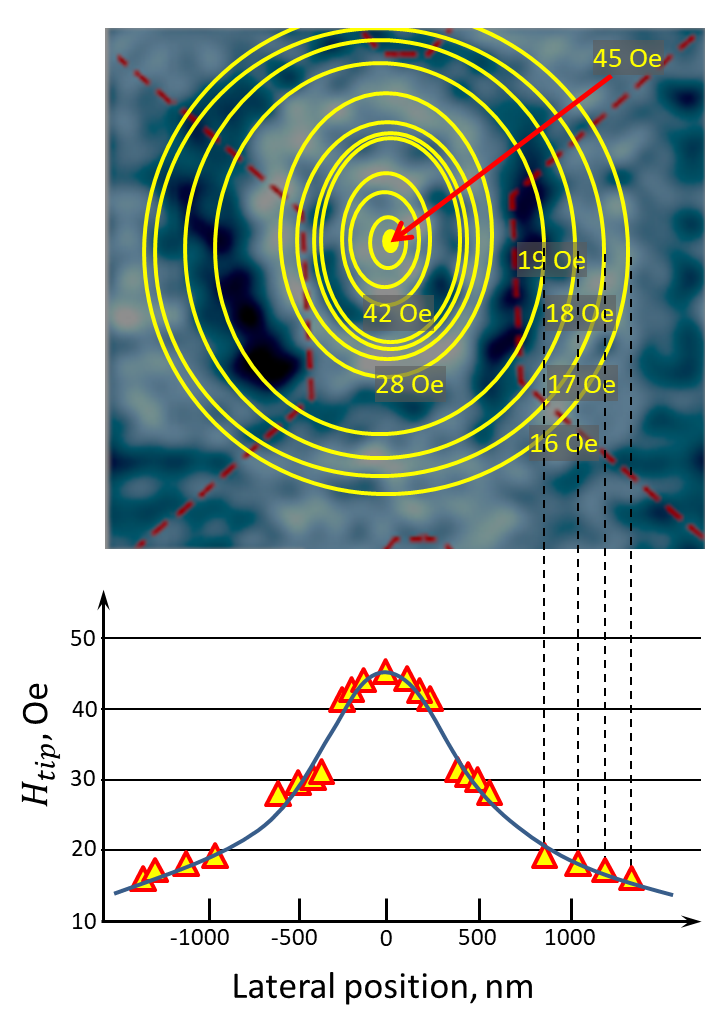}
        \caption{Extraction of tip field profile. Upper panel: constant tip field $H_{tip}(x,y)=const$ contours (yellow ellipses) extracted from the bright rings in Fig.4 following the second suggested calibration procedure. Lower panel: the cross-section of the extracted field profile in $y$-direction. Note that the field maximum is out of geometrical center of the LJJ (presented in background). 
        }
        \label{figS2}
    \end{center}
\end{figure}

If one needs to know the field profile closer to the magnetic charge, he/she needs to approach the tip closer to LJJ and to adjust the external field in order to compensate the increased magnetic flux. In this way, it is always possible to keep a well-determined number of Josephson vortices and use this fact for calibration.

If strongly magnetized probes are required (it is not usually the case as such MFM probes are invasive), their calibration still can be done by using LJJ of smaller lateral dimensions. Indeed, such LJJ are less sensitive to the magnetic field ($\Phi=H\times S$), thus shifting the range of Josephson vortex penetration to higher fields.

\section{Limitations of the proposed calibration approach} 

Due to flux quantization, the field intensity at which Josephson vortices penetrate depends mainly on the LJJ geometry, and not on transport current $I$ or temperature $T$. 
The amplitude $\Delta V$ of the effect does depend on  $I$ and $T$, but not the value of the magnetic field at which Josephson vortices penetrate the junction (see 3 experimental curves in Fig.2b).

While we demonstrated the MFM tip calibration, we can anticipate several factors limiting its use.

\begin{figure*}[!ht]
    \begin{center}
        \includegraphics[width=1.99\columnwidth]{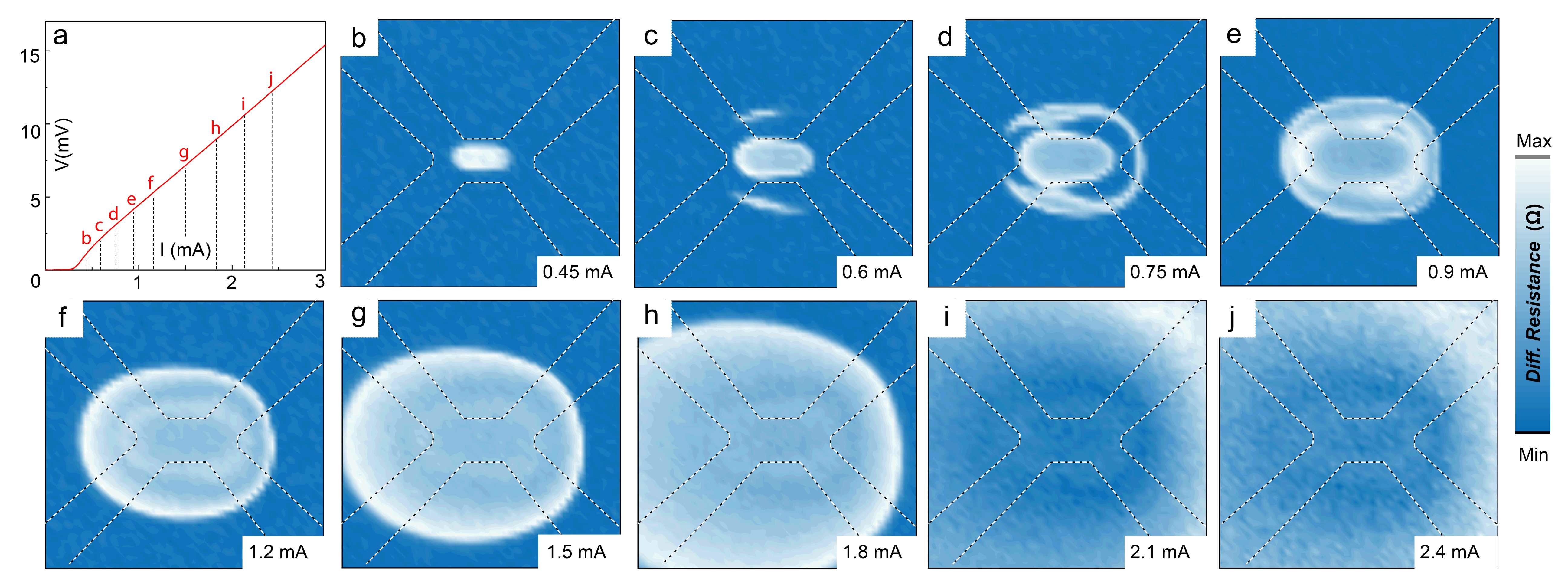}
        \caption{(a) $V(I)$ characteristics of LJJ with the cantilever placed close to the center $(x=0;y=0)$ at $h$ = 700 nm. (b-j) $7 \times 7 \mu$m$^2$ differential resistance maps $dV/dI(x,y)$ at different bias currents acquired with the cantilever scanned at $h$ = 700 nm above LJJ.} 
        \label{figS3}
    \end{center}
\end{figure*}

First, the temperature range is limited by the critical temperature of superconducting transition, $T_c$ = 7.4 K (see supplementary information in Ref.~\cite{dremov2019local}). Moreover, at the temperatures $T\lesssim T_c$ the total sensitivity of the method to Josephson vortex generation decreases thus reducing its precision.  

The second factor to consider is the effect of field focusing that we took into account via constant geometric factor $\alpha$ in Appendix B. This assumption is reasonable in a first approximation~\cite{stolyarov2020josephson,golod2021reconfigurable}. In general however, the spatially inhomogeneous field of the tip leads to a more complex distribution of Meissner currents  in LJJ and, as a consequence of the non-local current-field relation, to a non-trivial screening magnetic field focusing. Taking into account these effects is possible, but it requires more complex calculations of the field distribution using self-consistent Ginzburg-Landau approach, for example. 

Third, high external fields can influence the magnetic charge of the cantilever and/or Junction properties. At some kOe or higher, the tip magnetization changes, and even a charge sign inversion may occur.  But such high fields/magnetic moments are rarely used as they are too invasive for MFM imaging.

Finally, high external fields also alter the junction properties. Strong magnetic fields exceeding $H_{c1}$ of used superconductor lead to the penetration of Abrikosov vortices into superconducting leads of the junction and modify its response to the magnetic field. In our experiments in which Nb leads were used, $H_{c1} \sim$ 100 Oe. 
\end{document}